\documentclass[aps,pra,showpacs,twocolumn,superscriptaddress,amsmath,amssymb,nofootinbib]{revtex4}

\usepackage{graphicx,dcolumn,bm}
\usepackage{amsmath}
\usepackage{amsthm}
\usepackage{amsfonts}
\newcommand{\bra}[1]{\left\langle #1\right|}
\newcommand{\ket}[1]{\left|#1\right\rangle}
\newcommand{\braket}[2]{\left\langle #1\right|\left.#2\right\rangle}

\begin{document}

\title{Simulating the coupling of angular momenta in distant matter qubits} 

\author{C. Ammon}
\affiliation{Institut f\"ur Optik, Information und Photonik, Universit\"at Erlangen-N\"urnberg, D-91058 Erlangen, Germany}

\author{A. Maser}
\affiliation{Institut f\"ur Optik, Information und Photonik, Universit\"at Erlangen-N\"urnberg, D-91058 Erlangen, Germany}
\affiliation{Erlangen Graduate School in Advanced Optical Technologies (SAOT), Universit\"at Erlangen-N\"urnberg, 91052 Erlangen, Germany}

\author{U. Schilling}
\affiliation{Institut f\"ur Optik, Information und Photonik, Universit\"at Erlangen-N\"urnberg, D-91058 Erlangen, Germany}
\affiliation{Erlangen Graduate School in Advanced Optical Technologies (SAOT), Universit\"at Erlangen-N\"urnberg, 91052 Erlangen, Germany}

\author{T. Bastin}
\affiliation{Institut de Physique Nucl\'eaire, Atomique et de Spectroscopie, Universit\'e de Li\`ege, 4000 Li\`ege, Belgium}

\author{J. von Zanthier}
\email{joachim.vonzanthier@physik.uni-erlangen.de}
\homepage{http://www.optik.uni-erlangen.de/jvz/}
\affiliation{Institut f\"ur Optik, Information und Photonik, Universit\"at Erlangen-N\"urnberg, D-91058 Erlangen, Germany}
\affiliation{Erlangen Graduate School in Advanced Optical Technologies (SAOT), Universit\"at Erlangen-N\"urnberg, 91052 Erlangen, Germany}

\date{\today}

\begin{abstract}
We present a mathematical proof of the algorithm allowing to generate all - symmetric and non-symmetric - total angular momentum eigenstates in remote matter qubits by projective measurements, proposed in Maser \textit{et al.} [Phys.~Rev.~A \textbf{79}, 033833 (2009)]. By deriving a recursion formula for the algorithm we show that the generated states are equal to the total angular momentum eigenstates obtained via the usual quantum mechanical coupling of angular momenta. In this way we demonstrate that the algorithm is able to simulate the coupling of $N$ spin-1/2 systems, and to implement the required Clebsch-Gordan coefficients, even though the particles never directly interact with each other.
\end{abstract}

\pacs{42.50.Dv,42.50.Tx,37.10.-x,03.67.-a} 

\maketitle

\section{Introduction}

The generation of multipartite entangled states relies typically on a direct interaction between the particles, induced via nonlinear effects \cite{Shih88,Kwiat95}, collisions \cite{Osnaghi01}, Coulomb coupling \cite{Leibfried05, Haeffner05,Blatt11} or atom-photon interfaces \cite{Wilk07}. However, recently it was shown that 
it is also possible to entangle distant particles that never directly couple to each other. Examples are entanglement swapping of entangled photons \cite{Zukowski93} produced via Spontaneous Parametric Down Conversion (SPDC) \cite{Shih88,Kwiat95} or entanglement of remote atoms generated via projective measurements of emitted photons \cite{Julsgaard01, Chou05,Matsukevich06,Moehring07,Thiel07,Bastin09,Maser09}. By use of the latter technique, distant clouds of atoms have been entangled so far \cite{Julsgaard01, Chou05,Matsukevich06}, but it was also possible to demonstrate the entanglement of distant single atoms, e. g., two ions stored in different ion traps one meter apart \cite{Moehring07}. 

Recently, an algorithm was proposed, based on the technique of projective measurements, which allows the generation of all - symmetric and non-symmetric - total angular momentum eigenstates in $N$ distant matter qubits \cite{Maser09}, most of them being entangled states \cite{Dicke,Mandel}. The algorithm was motivated by comparison with the familiar construction in quantum mechanics of the eigenstates of a total angular momentum, obtained via coupling of $N$ spin-1/2 systems, and tested numerically for up to $N=8$ qubits. 

In this paper, we present a rigorous mathematical proof of this algorithm by the technique of induction.
For the proof, we derive two recursion formulas, one for the algorithm presented in \cite{Maser09} and the other one for the familiar quantum mechanical procedure to generate the eigenstates of a total angular momentum of an $N$-qubit compound. By complete induction it is then shown that both recursion formulas lead to the same result. In this way, we prove  that the algorithm presented in \cite{Maser09} allows to generate all symmetric and non-symmetric total angular momentum eigenstates, and to implement the required Clebsch-Gordan coefficients, for any number $N$ of spin-1/2 particles even though the particles never directly interact with each other. The algorithm is thus able to simulate the coupling of angular momenta even of non-interacting qubits.

The paper is organized as follows: in Secs.~II and~III we  recall the algorithm presented in \cite{Maser09} and formulate it in a rigorous mathematical way. In Sec.~IV we exemplify the algorithm for a specific case and, in Sec.~V, we rewrite it in form of a recursion relation. In Sect.~VI we recapitulate the well-known quantum mechanical rules to generate the eigenstates of a total angular momentum from the eigenstates of the angular momenta to be coupled and rewrite it for the special case of $N$ spin-1/2 particles also in form of a recursion formula. In Sec.~VII the two recursion formulas of Sec.~V and VI are then compared and proven to be identical by the technique of induction. In Sec.~VIII we finally conclude.

\section{Algorithm for the generation of total angular momentum eigenstates of $N$ spin-1/2 particles}

For $N$ spin-1/2 particles, the total angular momentum eigenstates, defined as simultaneous eigenstates of the square of the total spin operator $\hat{\bf S}^2_N$ and its $z$-component $\hat{S}_{N,z}$, are commonly denoted by $\ket{S_N;m_N}$, with the corresponding eigenvalues $S_N(S_N+1)\hbar^2$ and $m_N\hbar$, respectively \cite{Dicke,Mandel}. However, the denomination $\ket{S_N;m_N}$ may in general characterize more than one quantum state. We will thus specify more precisely an $N$ qubit total angular momentum eigenstate by incorporating in its notation its coupling history, i.e., by writing the quantum numbers $S_1, S_2, ..., S_{N-1}$ of the total spins ${\bf S}_1, {\bf S}_2, ..., {\bf S}_N$ in addition to $S_N$ and $m_N$. According to the quantum mechanical rules of adding angular momenta, a single qubit state has $S_1=\frac{1}{2}$, a two qubit system can either have $S_2=0$ or $S_2=1$, a three qubit system $S_3=\frac{1}{2}$ or $S_3=\frac{3}{2}$, and so on. Including the coupling history we thus get the notation $\ket{S_1,S_2,...,S_N;m_N}$ for an eigenstate of the total spin ${\bf S}_N$ which describes a particular angular momentum eigenstate unambiguously \cite{Maser09}.

To generate all total angular momentum eigenstates in non-interacting matter qubits, Maser {\textit et al.} considered in \cite{Maser09} $N$ uncorrelated single photon emitters with a $\Lambda$-configuration, e.g., trapped neutral atoms, trapped ions, quantum dots, or any other equivalent physical system with access to a similar behavior. In the beginning each emitter is assumed to be in the excited state $\ket{e}$. Subsequently, due to spontaneous decay, the emitters leave the excited state $\ket{e}$ and populate the two stable ground states $\ket{-}$ and $\ket{+}$ under the emission of a ${\bm \sigma^+}$ and ${\bm \sigma^-}$ polarized photon, respectively; hereby, the two states $\ket{\pm}$ encode the qubit, with eigenvalues $\pm \hbar/2$ for the operator $\hat{S}_{1,z}$. The $N$ photons are either registered in the far field by $N$ single photon detectors or collected and transmitted via optical single mode fibers towards the $N$ detectors where it is assumed that each detector registers exactly one photon. Hereby, each photon can possibly travel via $N$ paths towards the $N$ detectors so that after a successful detection event it is impossible to determine along which way a recorded photon has travelled. This causes multi photon intereference effects of $N$th order which can be exploited to generate a large variety of entangled multi qubit states \cite{Thiel07, Bastin09, Maser09}.

Using this scheme and assuming that optical fibers for the transmission of the photons towards the $N$ detectors are used, each accumulating an optical phase $\delta_{j,k}= 2\pi n$ (with $n\in\mathbb{Z}$) between emitter $k$ and detector $j$, the algorithm for the generation of any of the $2^N$ symmetric and nonsymmetric total angular momentum eigenstates $\ket{S_1,S_2,S_3,...,S_N;m_N}$ spanning the Hilbert space of the $N$-qubit compound can be formulated in the following way \cite{Maser09}:  

\begin{itemize}
\item[1.] set up $\frac{N}{2}+m_N$ ($\frac{N}{2}-m_N$) detectors with $\boldsymbol{\sigma}^-$ ($\boldsymbol{\sigma}^+$) polarized filters in front. Hereby, we connect the first emitter with optical fibers to all $N$ detectors.
\item[2.] check for each emitter $k$ beginning with $k=2$ whether $S_k>S_{k-1}$ or $S_k<S_{k-1}$. If
\begin{itemize}
\item[a.] $S_k>S_{k-1}$, we have to connect the emitter with optical fibers to all detectors except those which are mentioned in case [b.] below.
\item[b.] $S_k<S_{k-1}$, we have to connect the emitter with optical fibers to one detector with a $\boldsymbol{\sigma}^-$ polarizer and to one with a $\boldsymbol{\sigma}^+$ polarizer where the optical fiber leading to the $\boldsymbol{\sigma}^-$ polarizer should induce an additional relative optical phase shift of $\pi$. Those two detectors should not be linked to any other subsequent emitter.
\end{itemize}
\end{itemize}
If one wants to create a particular total angular momentum eigenstate $|{S_1,S_2,S_3,...,S_N;m_N}\rangle$, the setup to generate this state is unambiguously determined by the quantum numbers $S_1,S_2,S_3,...,S_N$ obtained by successively coupling the spins of $N$ spin-1/2 particles, and the magnetic quantum number $m_N$ of the total spin $\textbf{S}_N$. Hereby, according to the rules of the algorithm presented above, $m_N$ determines the number of $\boldsymbol{\sigma}^-$ and $\boldsymbol{\sigma}^+$ polarising filters used in the setup.

\section{Mathematical formulation of the algorithm}

In order to prove the algorithm presented in \cite{Maser09} and recalled in Sect.~II, we have to formulate it in a more mathematical way. To this end, we note that if emitter $k$ emits a photon and we register this photon by any of the $N$ detectors $j$ ($j,k \in\{1,...,N\}$), which is however not precisely known, equipped with a polarising filter in front with orientation $\boldsymbol{\epsilon}_j=\alpha_j\boldsymbol{\sigma}^++\beta_j\boldsymbol{\sigma}^-$, the state of our atomic system is projected by applying the following projection operator $\hat{P}_k$ \cite{Thiel07, Bastin09}:
\begin{equation}
\label{modepop}
\hat{P}_k=\sum_{j=1}^N e^{i\delta_{j,k}} \left(\alpha_j\ket{-}_k\bra{e}+\beta_j \ket{+}_k\bra{e}\right)\otimes\ket{1}_j\bra{0}.
\end{equation}
Hereby, the phase factor $e^{i\delta_{j,k}}$ accounts for the optical phase $\delta_{j,k}$ accumulated on the way from emitter $k$ to detector $j$ and the operator $\ket{\pm}_k\bra{e}$ projects the $k$th atom from the excited state $\ket{e}$ to the ground state $\ket{\pm}$, while leaving the state of the other atoms unchanged. In Eq.~(\ref{modepop}) the operator $\ket{1}_j\bra{0}$ ensures that every detector registers exactly one photon. While this is not relevant when considering only the photon emitted by a single atom it becomes important when each of several atoms scatters one photon. Since it is assumed that initially all atoms are in the excited state and all of the photonic modes are unpopulated, the initial state $\ket{\psi_{i,N}}$ of the atoms and the modes is given by
\begin{equation}
\ket{\psi_{i,N}}=\ket{e...e}\otimes\ket{0...0},
\end{equation}

After recording all $N$ photons at the $N$ detectors, the final state $\ket{\psi_f}$ of the system can be written in the following form
\begin{widetext}
\begin{equation} 
\label{final_state}
\begin{split}
\ket{\psi_f}=\hat{P}_N\cdots &\hat{P}_1\ket{\psi_{i,N}}=\\
&\sum_{k=0}^N \frac{1}{k!(N-k)!}\sum_{\tau\in \Sigma_N}\sum_{\sigma\in \Sigma_N} \beta_{\sigma(1)}\cdots\beta_{\sigma(k)}\alpha_{\sigma(k+1)}\cdots\alpha_{\sigma(N)}\chi_{\sigma(1)\tau(1)}\cdots\chi_{\sigma(N)\tau(N)}\\
&\underbrace{\left(\ket{+}_{\tau(1)}\bra{e}\otimes\cdots\otimes\ket{+}_{\tau(k)}\bra{e}\otimes\ket{-}_{\tau(k+1)}\bra{e}\otimes\cdots\otimes\ket{-}_{\tau(N)}\bra{e}\right)}_{\mbox{projection operator part}}\ket{\psi_{i,N}}\\\\
\end{split}
\end{equation}
\end{widetext}
with $e^{i\delta_{\sigma(j),\tau(k)}}=\chi_{\sigma(j)\tau(k)}$ and $j,k\in\{1,...,N\}$. Hereby, the second and third sum run over all permutations $\sigma,\tau$ of the set $\{1,...,N\}$, which can be mathematically denoted by the elements of the symmetric group $\Sigma_N$ \cite{grouptheory}. The summation index $k$ of the first sum controls the number of $\beta-$ and $\alpha-$coefficients, i.e., the number of projection operators $\ket{+}\bra{e}$ and $\ket{-}\bra{e}$. Since we sum over all permutations of the projection operator for every $k\in\{0,...,N\}$, we have to compensate the multiple counts in the sum $\sum_{\tau\in \Sigma_N}$ with the factor $\frac{1}{k!(N-k)!}$: there are $k!$ combinations to choose the same set of projection operators $\ket{+}\bra{e}$ and $(N-k)!$ possibilities to choose the same set of projection operators $\ket{-}\bra{e}$. The sum $\sum_{\sigma\in \Sigma_N} \beta_{\sigma(1)}\cdots\beta_{\sigma(k)}\alpha_{\sigma(k+1)}\cdots\alpha_{\sigma(N)}$ must not be compensated for multiple counts, because these multiple counts are already included in the product $\hat{P}_N\cdots\hat{P}_1$, as each $\hat{P}_k$ is a sum over all $\alpha$'s and $\beta$'s (cf. Eq.~(\ref{modepop})).

In general, the phase factors $\chi_{jk}$ may take any complex value with norm between zero and one in order to allow for an attenuation of the light fields propagating from emitter $k$ to detector $j$. However, in the presented algorithm $\chi_{jk}$ needs only to take the values $1$ (``normal'', i.e., not attenuated connection between emitter and detector with $\delta_{j,k}=2\pi n$, where $n\in\mathbb{Z}$), $0$ (no connection between emitter and detector, meaning an attenuation of $100\%$) and $-1$ (not attenuated connection between emitter and detector with an additional phase shift of $\pi$ with respect to a normal connection).

Note that we omitted in Eq.~(\ref{final_state}) the operators $\ket{1}_j\bra{0}$. They were necessesary to derive Eq.~(\ref{final_state}) but would in the final form of Eq.~(\ref{final_state}) only add a global factor of $\prod^N_{j=0}\ket{1}_j\bra{0}$, which does not influence the final atomic state. Therefore, we chose to omit this factor for the sake of simplicity and accordingly change the initial state to
\begin{equation}
\ket{\psi_{i,N}}=\ket{e\ldots e}.
\end{equation}

\section{Exemplification of the algorithm}

\begin{figure}[h!]
\centering
\includegraphics[width=8cm]{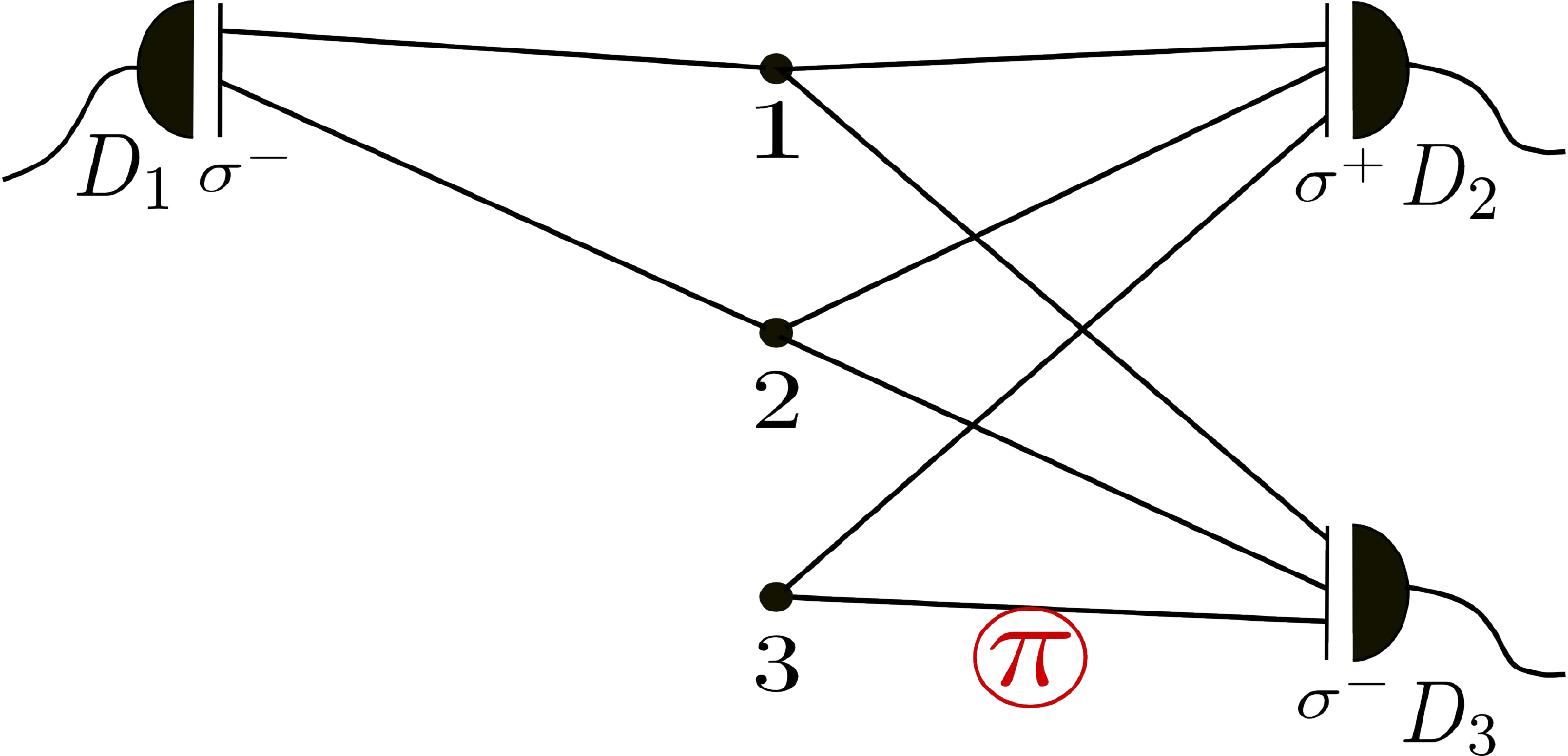}
\caption{\small\slshape{Setup for the generation of the state $\ket{\frac{1}{2},1,\frac{1}{2};\frac{1}{2}}$.}}
\label{Algo}
\end{figure}

As an example for the operation of the algorithm, let us generate the three qubit state $\ket{\frac{1}{2},1,\frac{1}{2};\frac{1}{2}}$, where we have $S_1 = 1/2$, $S_2= 1$, $S_3 = 1/2$ and $m_3 = 1/2$. This state can be written in the decoupled basis as $\frac{1}{\sqrt{6}}(2\ket{++-}-\ket{+-+}-\ket{-++})$ \cite{Dicke}. Following the prescriptions of the algorithm of Sect.~II we need $\frac{N}{2}+m_N=\frac{3}{2}+\frac{1}{2}=2$ $\boldsymbol{\sigma}^--$polarising filters in front of any two out of three detectors. As the index $k$ of the first sum in Eq.~(\ref{final_state}) corresponds to the number of polarisers with $\boldsymbol{\sigma}^--$orientation, the only non-zero contribution to the sum comes from the summand with $k=2$. If we use the well-known cycle notation from group theory \cite{grouptheory} for the elements of the symmetric group $\Sigma_3$, we can expand Eq.~(\ref{final_state}) in the following way:
\begin{widetext}
\begin{equation}
\label{eq:example}
\begin{split}
\hat{P}_3\hat{P}_2\hat{P}_1\ket{\psi_{i,3}}=\frac{1}{2}&\left(\underbrace{\beta_1\beta_2\alpha_3\chi_{11}\chi_{22}\chi_{33}}_{\sigma=id,\tau=id}+\underbrace{\beta_2\beta_1\alpha_3\chi_{21}\chi_{12}\chi_{33}}_{\sigma=(12),\tau=id}+\underbrace{\beta_3\beta_2\alpha_1\chi_{31}\chi_{22}\chi_{13}}_{\sigma=(13),\tau=id}\right.\\
&\left.+\underbrace{\beta_1\beta_3\alpha_2\chi_{11}\chi_{32}\chi_{23}}_{\sigma=(23),\tau=id}+\underbrace{\beta_2 \beta_3\alpha_1\chi_{21}\chi_{32}\chi_{13}}_{\sigma=(123),\tau=id}+\underbrace{\beta_3\beta_1\alpha_2\chi_{31}\chi_{12}\chi_{23}}_{\sigma=(132),\tau=id}\right)\ket{++-}\\
+\frac{1}{2}&\left(\underbrace{\beta_1\beta_2\alpha_3\chi_{12}\chi_{21}\chi_{33}}_{\sigma=id,\tau=(12)}+\underbrace{\beta_2\beta_1\alpha_3\chi_{22}\chi_{11}\chi_{33}}_{\sigma=(12),\tau=(12)}+\underbrace{\beta_3\beta_2\alpha_1\chi_{32}\chi_{21}\chi_{13}}_{\sigma=(13),\tau=(12)}\right.\\
&\left.+\underbrace{\beta_1\beta_3\alpha_2\chi_{12}\chi_{31}\chi_{23}}_{\sigma=(23),\tau=(12)}+\underbrace{\beta_2 \beta_3\alpha_1\chi_{22}\chi_{31}\chi_{13}}_{\sigma=(123),\tau=(12)}+\underbrace{\beta_3\beta_1\alpha_2\chi_{32}\chi_{11}\chi_{23}}_{\sigma=(132),\tau=(12)}\right)\ket{++-}\\
+\frac{1}{2}&\left(\underbrace{\beta_1\beta_2\alpha_3\chi_{13}\chi_{22}\chi_{31}}_{\sigma=id,\tau=(13)}+\underbrace{\beta_2\beta_1\alpha_3\chi_{23}\chi_{12}\chi_{31}}_{\sigma=(12),\tau=(13)}+\underbrace{\beta_3\beta_2\alpha_1\chi_{33}\chi_{22}\chi_{11}}_{\sigma=(13),\tau=(13)}\right.\\
&\left.+\underbrace{\beta_1\beta_3\alpha_2\chi_{13}\chi_{32}\chi_{21}}_{\sigma=(23),\tau=(13)}+\underbrace{\beta_2 \beta_3\alpha_1\chi_{23}\chi_{32}\chi_{11}}_{\sigma=(123),\tau=(13)}+\underbrace{\beta_3\beta_1\alpha_2\chi_{33}\chi_{12}\chi_{21}}_{\sigma=(132),\tau=(13)}\right)\ket{-++}\\
+\frac{1}{2}&\left(\underbrace{\beta_1\beta_2\alpha_3\chi_{11}\chi_{23}\chi_{32}}_{\sigma=id,\tau=(23)}+\underbrace{\beta_2\beta_1\alpha_3\chi_{21}\chi_{13}\chi_{32}}_{\sigma=(12),\tau=(23)}+\underbrace{\beta_3\beta_2\alpha_1\chi_{31}\chi_{23}\chi_{12}}_{\sigma=(13),\tau=(23)}\right.\\
\end{split}
\end{equation} 
\begin{equation*}
\begin{split}
&\left.+\underbrace{\beta_1\beta_3\alpha_2\chi_{11}\chi_{33}\chi_{22}}_{\sigma=(23),\tau=(23)}+\underbrace{\beta_2 \beta_3\alpha_1\chi_{21}\chi_{33}\chi_{12}}_{\sigma=(123),\tau=(23)}+\underbrace{\beta_3\beta_1\alpha_2\chi_{31}\chi_{13}\chi_{22}}_{\sigma=(132),\tau=(23)}\right)\ket{+-+}\\+\frac{1}{2}&\left(\underbrace{\beta_1\beta_2\alpha_3\chi_{12}\chi_{23}\chi_{31}}_{\sigma=id,\tau=(123)}+\underbrace{\beta_2\beta_1\alpha_3\chi_{22}\chi_{13}\chi_{31}}_{\sigma=(12),\tau=(123)}+\underbrace{\beta_3\beta_2\alpha_1\chi_{32}\chi_{23}\chi_{11}}_{\sigma=(13),\tau=(123)}\right.\\
&\left.+\underbrace{\beta_1\beta_3\alpha_2\chi_{12}\chi_{33}\chi_{21}}_{\sigma=(23),\tau=(123)}+\underbrace{\beta_2 \beta_3\alpha_1\chi_{22}\chi_{33}\chi_{11}}_{\sigma=(123),\tau=(123)}+\underbrace{\beta_3\beta_1\alpha_2\chi_{32}\chi_{13}\chi_{21}}_{\sigma=(132),\tau=(123)}\right)\ket{-++}\\
+\frac{1}{2}&\left(\underbrace{\beta_1\beta_2\alpha_3\chi_{13}\chi_{21}\chi_{32}}_{\sigma=id,\tau=(132)}+\underbrace{\beta_2\beta_1\alpha_3\chi_{23}\chi_{11}\chi_{32}}_{\sigma=(12),\tau=(132)}+\underbrace{\beta_3\beta_2\alpha_1\chi_{33}\chi_{21}\chi_{12}}_{\sigma=(13),\tau=(132)}\right.\\
&\left.+\underbrace{\beta_1\beta_3\alpha_2\chi_{13}\chi_{31}\chi_{22}}_{\sigma=(23),\tau=(132)}+\underbrace{\beta_2 \beta_3\alpha_1\chi_{23}\chi_{31}\chi_{12}}_{\sigma=(123),\tau=(132)}+\underbrace{\beta_3\beta_1\alpha_2\chi_{33}\chi_{11}\chi_{22}}_{\sigma=(132),\tau=(132)}\right)\ket{+-+}\\
\end{split}
\end{equation*} 
\end{widetext} 
\quad\\
\noindent According to the recipe of the algorithm we could set up the polarization filters and connect the emitters to the detectors for example according to the scheme depicted in Fig.~\ref{Algo}.

From the figure, we can now determine the required values for the coefficients $\alpha_j,\beta_j$ and the phase factors $\chi_{jk}$ appearing in Eq.~(\ref{eq:example}). 
There are two ${\bm \sigma^--}$ polarising filters, one in front of detector $1$ and the other one in front of detector $3$, thus we have 
\begin{equation}
\label{eq:coeff1}
\alpha_1=\alpha_3=0\quad \wedge \quad \beta_1=\beta_3=1.	
\end{equation}
In front of detector $2$, a ${\bm \sigma^+-}$polarising filter is positioned and therefore we have to set
\begin{equation}
\alpha_2=1\quad \wedge \quad \beta_2=0.	
\end{equation}
A normal connection between an emitter $k$ and a detector $j$ has an optical phase factor $\chi_{jk}=1$ (depicted in Fig.\ref{Algo} by a black line). A missing link indicates an attenuation of $100\%$ and thus $\chi_{jk}=0$. Last, the optical fiber loop (labeled in Fig.\ref{Algo} with $\pi$) indicates an additional optical phase shift of $\pi$ and hence $\chi_{jk}=-1$. In short we thus get: 
\begin{equation}
\begin{split}
\label{eq:coeff3}
\quad \chi_{11}=\chi_{21}=&\chi_{31}=\chi_{12}=\chi_{22}=\chi_{32}=\chi_{23}=1 \\
\quad &\chi_{33}=-1 \quad \wedge \quad \chi_{13}=0
\end{split}
\end{equation}
Hence, we can calculate the final state by inserting Eq.~(\ref{eq:coeff1})-(\ref{eq:coeff3}) into Eq.~(\ref{eq:example}), what leads to:
\begin{equation}
\label{tz}
\left.\hat{P}_3\hat{P}_2\hat{P}_1\ket{\psi_{i,3}}\right|_{\ket{\frac{1}{2},1,\frac{1}{2};\frac{1}{2}}}=2\ket{++-}-\ket{+-+}-\ket{-++}.
\end{equation} 
corresponding - up to a normalization constant - to the desired state $\ket{\frac{1}{2},1,\frac{1}{2};\frac{1}{2}}$. 

Note that we introduced in Eq.~(\ref{tz}) a new notation when writing the state as $\left.\hat{P}_N\cdots\hat{P}_1\ket{\psi_{i,N}}\right|_{\ket{S_1,...,S_N;m_N}}$. Hereby, we took into account that the quantum numbers $S_1,S_2,...,S_N$, labeling the state $\ket{S_1,...,S_N;m_N}$, determine	unambiguously the values for $\alpha_j$, $\beta_j$ and $\chi_{jk}$ (with $j,k\in\{1,...,N\}$). Thus, in order to describe exactly the state we want to create we have to include these quantum numbers to the state on the left hand side of Eq.~(\ref{final_state}). This allows us to know exactly the particular values we have to choose for $\alpha_j$, $\beta_j$ and $\chi_{jk}$ in order to implement the targeted state $\ket{S_1,...,S_N;m_N}$ by use of the algorithm. Later on, we will also use a notation of the form $\left.\hat{P}_N\cdots\hat{P}_1\ket{\psi_{i,N}}\right|_{k=x}$, which means that the sum in Eq.~(\ref{final_state}) over the index $k$ reduces to the summand with $k=x$.\\

\section{Recursivity of the algorithm}

Maser \textit{et al.} have shown in \cite{Maser09} that the generation of the state $\ket{\frac{1}{2},1,\frac{1}{2};\frac{1}{2}}$ following the recipe of the algorithm is in fact obtained in two steps. First, the states $\ket{\frac{1}{2},1;1}$ and $\ket{\frac{1}{2},1;0}$ are generated and then, in a second step, these two states are superposed by use of another qubit $\ket{\frac{1}{2};\pm \frac{1}{2}}$, just like quantum mechanically coupling a spin-1 system with another spin-1/2 system \cite{Maser09}.

Inspired by this procedure, we will try to retrieve a general recursivity in the mathematical formulation of our algorithm given in Eq.~(\ref{final_state}). This means that we have to find a recursion formula for $\left.\hat{P}_N\cdots\hat{P}_1\ket{\psi_{i,N}}\right|_{\ket{S_1,...,S_N;m_N}}$.

As shown in the previous example, only the $k^\prime$th summand in Eq.~(\ref{final_state}) is unequal to zero, because the algorithm tells us to set up exactly $k^\prime=\frac{N}{2}+m_N$ $\boldsymbol{\sigma}^-$ polarizers (what means that exactly $k^\prime$ $\beta$-coefficients are unequal zero and therefore the sum over $k$ in Eq.~(\ref{final_state}) reduces to $k=k^\prime$).

In the following we will divide Eq.~(\ref{final_state}) into two sums. The first sum corresponds to Eq.~(\ref{final_state}) for $N-1$ atoms times the projection of the $N$th atom to the state $\ket{+}$ and the second sum corresponds to Eq.~(\ref{final_state}) for $N-1$ atoms times the projection of the $N$th atom to the state $\ket{-}$. Hereby, for the $N$th atom, we can choose any of the $\left(\frac{N}{2} \pm m_N\right)$ 
atoms which are projected into the state $\ket{\pm}$ 
, thus we have to consider an additional factor $\left(\frac{N}{2} \pm m_N\right)$ 
for both sums.

This decomposition can again be divided into $N$ sums, since in principal every detector can project the $N$th atom into the desired state. After this partitioning, we can exclude the factor $\beta_i\chi_{iN}$ ($\alpha_i\chi_{iN}$) what means that the $N$th atom is projected into the state $\ket{+}$ ($\ket{-}$) by registering the emitted photon at detector $i$. Thus, altogether we arrive at

\begin{widetext}
\begin{displaymath}
\begin{split}
&\left.\hat{P}_N\cdots\hat{P}_1\ket{\psi_{i,N}}\right|_{k=\left(\frac{N}{2}+m_N\right)}=\frac{1}{\left(\frac{N}{2}+m_N\right)!\left(\frac{N}{2}-m_N\right)!}\\
&\sum_{\tau\in \Sigma_N}\sum_{\sigma\in \Sigma_N} \beta_{\sigma(1)}\cdots\beta_{\sigma\left(\frac{N}{2}+m_N\right)}\alpha_{\sigma\left(\frac{N}{2}+m_N+1\right)}\cdots\alpha_{\sigma(N)}\chi_{\sigma(1)\tau(1)}\cdots\chi_{\sigma(N)\tau(N)}\\
&\left(\ket{+}_{\tau(1)}\bra{e}\otimes\cdots\otimes\ket{+}_{\tau\left(\frac{N}{2}+m_N\right)}\bra{e}\otimes\ket{-}_{\tau\left(\frac{N}{2}+m_N+1\right)}\bra{e}\otimes\cdots\otimes\ket{-}_{\tau(N)}\bra{e}\right)\ket{\psi_{i,N}}\\
\end{split}
\end{displaymath}
\begin{displaymath}
\begin{split}
&=\frac{\left(\frac{N}{2}+m_N\right)}{\left(\frac{N}{2}+m_N\right)!\left(\frac{N}{2}-m_N\right)!}\\
&\left(\beta_{1}\chi_{1N}\sum_{\tau\in \Sigma_{N-1}}\sum_{\sigma\in \Sigma_{N\setminus(1)}}\beta_{\sigma(2)} \cdots\beta_{\sigma\left(\frac{N}{2}+m_N\right)}\alpha_{\sigma\left(\frac{N}{2}+m_N+1\right)}\cdots\alpha_{\sigma(N)}\chi_{\sigma(2)\tau(1)}\cdots\right.\\
&\hspace{2cm}\left.\cdots\chi_{\sigma(N)\tau(N-1)}\left(\ket{+}_{\tau(1)}\bra{e}\otimes\cdots
\otimes\ket{+}_{\tau\left(\frac{N}{2}+m_N-1\right)}\bra{e}\otimes\ket{-}_{\tau\left(\frac{N}{2}+m_N\right)}\bra{e}\right.\right.\\
&\hspace{5cm}\left.\left.\otimes\cdots\otimes\ket{-}_{\tau(N-1)}\bra{e}\otimes\ket{+}_N\bra{e}\right)\ket{\psi_{i,N}}\right.\\
&\hspace{2cm}\vdots\\
\end{split}
\end{displaymath}
\begin{equation}
\label{beta}
\begin{split}
&\hspace{0.25cm}\left.+\beta_{N}\chi_{NN}\sum_{\tau\in \Sigma_{N-1}}\sum_{\sigma\in \Sigma_{N\setminus(N)}}\beta_{\sigma(1)} \cdots\beta_{\sigma\left(\frac{N}{2}+m_N-1\right)}\alpha_{\sigma\left(\frac{N}{2}+m_N\right)}\cdots\alpha_{\sigma(N-1)}\chi_{\sigma(1)\tau(1)}\cdots\right.\\
&\hspace{2cm}\left.\cdots\chi_{\sigma(N-1)\tau(N-1)}\left(\ket{+}_{\tau(1)}\bra{e}\otimes\cdots\otimes\ket{+}_{\tau\left(\frac{N}{2}+m_N-1\right)}\bra{e}\otimes\ket{-}_{\tau\left(\frac{N}{2}+m_N\right)}\bra{e}\right.\right.\\
&\hspace{5.25cm}\left.\left.\otimes\cdots\otimes\ket{-}_{\tau(N-1)}\bra{e}\otimes\ket{+}_N\bra{e}\right)\ket{\psi_{i,N}}\vphantom{\sum_1^N}\right)\\
\end{split}
\end{equation}
\begin{equation}
\label{alpha}
\begin{split}
&+\frac{\left(\frac{N}{2}-m_N\right)}{\left(\frac{N}{2}+m_N\right)!\left(\frac{N}{2}-m_N\right)!}\\
&\left(\alpha_{1}\chi_{1N}\sum_{\tau\in \Sigma_{N-1}}\sum_{\sigma\in \Sigma_{N\setminus(1)}}\beta_{\sigma(2)} \cdots\beta_{\sigma\left(\frac{N}{2}+m_N+1\right)}\alpha_{\sigma\left(\frac{N}{2}+m_N+2\right)}\cdots\alpha_{\sigma(N)}\chi_{\sigma(2)\tau(1)}\cdots\right.\\
&\hspace{2cm}\left.\cdots\chi_{\sigma(N)\tau(N-1)}\left(\ket{+}_{\tau(1)}\bra{e}\otimes\cdots\otimes\ket{+}_{\tau\left(\frac{N}{2}+m_N\right)}\bra{e}\otimes\ket{-}_{\tau\left(\frac{N}{2}+m_N+1\right)}\bra{e}\right.\right.\\
&\hspace{5cm}\left.\left.\otimes\cdots\otimes\ket{-}_{\tau(N-1)}\bra{e}\otimes\ket{-}_N\bra{e}\right)\ket{\psi_{i,N}}\right.\\
&\hspace{2cm}\vdots\\
&\hspace{0.25cm}\left.+\alpha_{N}\chi_{NN}\sum_{\tau\in \Sigma_{N-1}}\sum_{\sigma\in \Sigma_{N\setminus(N)}}\beta_{\sigma(1)} \cdots\beta_{\sigma\left(\frac{N}{2}+m_N\right)}\alpha_{\sigma\left(\frac{N}{2}+m_N+1\right)}\cdots\alpha_{\sigma(N-1)}\chi_{\sigma(1)\tau(1)}\cdots\right.\\
&\hspace{2cm}\left.\cdots\chi_{\sigma(N-1)\tau(N-1)}\left(\ket{+}_{\tau(1)}\bra{e}\otimes\cdots\otimes\ket{+}_{\tau\left(\frac{N}{2}+m_N\right)}\bra{e}\otimes\ket{-}_{\tau\left(\frac{N}{2}+m_N+1\right)}\bra{e}\right.\right.\\
&\hspace{5.25cm}\left.\left.\otimes\cdots\otimes\ket{-}_{\tau(N-1)}\bra{e}\otimes\ket{-}_N\bra{e}\right)\ket{\psi_{i,N}}\vphantom{\sum_1^N}\right).\\
\end{split}
\end{equation}
\end{widetext}
Hereby, we denoted with $\Sigma_{N\setminus(i)}$ in Eqs.~(\ref{beta}) and (\ref{alpha}) the symmetric group of the set $\{1,...,N\}\setminus\{i\}$ and with $\Sigma_{N-1}$ the symmetric group of the set \\$\{1,...,N-1\}$. Before simplifying Eqs.~(\ref{beta}) and (\ref{alpha}) further, we want to prove the following identity:

\begin{widetext}

When applying our algorithm in the mathematical formulation of Sect.~III (Eq.~(\ref{final_state})) to the generation of an arbitrary $N-1$ qubit total angular momentum eigenstate in an $(N-1)$-qubit compound system , the following equation holds:
\begin{equation}
\label{lemma}
\begin{split}
&\sum_{\sigma\in \Sigma_{N\setminus(1)}}\beta_{\sigma(2)} \cdots\beta_{\sigma\left(k\right)}\alpha_{\sigma\left(k+1\right)}\cdots\alpha_{\sigma(N-1)}\alpha_{\sigma(N)}\chi_{\sigma(2)\tau(1)}\cdots\chi_{\sigma(N)\tau(N-1)}\\
=&\sum_{\sigma\in \Sigma_{N\setminus(2)}}\beta_{\sigma(1)}\beta_{\sigma(3)} \cdots\beta_{\sigma\left(k\right)}\alpha_{\sigma\left(k+1\right)}\cdots\alpha_{\sigma(N-1)}\alpha_{\sigma(N)}\chi_{\sigma(1)\tau(1)}\cdots\chi_{\sigma(N)\tau(N-1)}\\
&\hspace{2cm}\vdots\\
=&\sum_{\sigma\in \Sigma_{N\setminus(N)}}\beta_{\sigma(1)}\cdots\beta_{\sigma\left(k-1\right)}\alpha_{\sigma\left(k\right)}\cdots\alpha_{\sigma(N-1)}\chi_{\sigma(1)\tau(1)}\cdots\chi_{\sigma(N-1)\tau(N-1)}.\\
\end{split}
\end{equation} 
\end{widetext}

The identity follows from the prescription of the algorithm and the corresponding degrees of freedom: according to the algorithm, 
for the generation of the total angular momentum eigenstates for $N-1$ qubits, we can freely choose any $N-1$ out of the $N$ detectors for the required setup. We exploit this degree of freedom by choosing a different detector configuration in each line of Eq.(\ref{lemma}). In this way, we can substitute the $N$th detector with any detector $i\in\{1,...,N-1\}$.

After that substitution we can also write
\begin{widetext}
\begin{displaymath}
\begin{split}
&\sum_{\sigma\in \Sigma_{N-1}}\beta_{\sigma(2)} \cdots\beta_{\sigma\left(k\right)}\alpha_{\sigma\left(k+1\right)}\cdots\alpha_{\sigma(N-1)}\alpha_{\sigma(1)}\chi_{\sigma(2)\tau(1)}\cdots\chi_{\sigma(1)\tau(N-1)}\\
=&\sum_{\sigma\in \Sigma_{N-1}}\beta_{\sigma(1)}\beta_{\sigma(3)} \cdots\beta_{\sigma\left(k\right)}\alpha_{\sigma\left(k+1\right)}\cdots\alpha_{\sigma(N-1)}\alpha_{\sigma(2)}\chi_{\sigma(2)\tau(1)}\cdots\chi_{\sigma(2)\tau(N-1)}\\
&\hspace{2cm}\vdots\\
=&\sum_{\sigma\in \Sigma_{N-1}}\beta_{\sigma(1)}\cdots\beta_{\sigma\left(k-1\right)}\alpha_{\sigma\left(k\right)}\cdots\alpha_{\sigma(N-1)}\chi_{\sigma(1)\tau(1)}\cdots\chi_{\sigma(N-1)\tau(N-1)}\\
\end{split}
\end{displaymath}
\end{widetext}
what proves Eq.~(\ref{lemma}).\\

By using Eq.~(\ref{lemma}) in Eq.~(\ref{beta}) (with $k=\frac{N}{2}+m_N$) and in Eq.~(\ref{alpha}) (with $k=\frac{N}{2}+m_N+1$) we can simplify Eqs.~(\ref{beta}) and (\ref{alpha}) to: 

\begin{widetext}
\begin{displaymath}
\begin{split}
\allowdisplaybreaks
&\left.\hat{P}_N\cdots\hat{P}_1\ket{\psi_{i,N}}\right|_{k=\left(\frac{N}{2}+m_N\right)}=\frac{1}{\left(\frac{N}{2}+m_N-1\right)!\left(\frac{N}{2}-m_N\right)!}\\
&\left(\sum_{i=1}^N\beta_{i}\chi_{iN}\right)\sum_{\tau\in \Sigma_{N-1}}\sum_{\sigma\in \Sigma_{N-1}}\beta_{\sigma(1)} \cdots\beta_{\sigma\left(\frac{N}{2}+m_N-1\right)}\alpha_{\sigma\left(\frac{N}{2}+m_N\right)}\cdots\alpha_{\sigma(N-1)}\chi_{\sigma(1)\tau(1)}\cdots\\
&\hspace{2cm}\cdots\chi_{\sigma(N-1)\tau(N-1)}\left(\ket{+}_{\tau(1)}\bra{e}\otimes\cdots\otimes\ket{+}_{\tau\left(\frac{N}{2}+m_N-1\right)}\bra{e}\otimes\ket{-}_{\tau\left(\frac{N}{2}+m_N\right)}\bra{e}\right.\\
&\hspace{6cm}\left.\otimes\cdots\otimes\ket{-}_{\tau(N-1)}\bra{e}\otimes\ket{+}_N\bra{e}\right)\ket{\psi_{i,N}}\\
&\hspace{0.25cm}+\frac{1}{\left(\frac{N}{2}+m_N\right)!\left(\frac{N}{2}-m_N-1\right)!}\\
&\left(\sum_{i=1}^N\alpha_{i}\chi_{iN}\right)\sum_{\tau\in \Sigma_{N-1}}\sum_{\sigma\in \Sigma_{N-1}}\beta_{\sigma(1)} \cdots\beta_{\sigma\left(\frac{N}{2}+m_N\right)}\alpha_{\sigma\left(\frac{N}{2}+m_N+1\right)}\cdots\alpha_{\sigma(N-1)}\chi_{\sigma(1)\tau(1)}\cdots\\
&\hspace{2cm}\cdots\chi_{\sigma(N-1)\tau(N-1)}\left(\ket{+}_{\tau(1)}\bra{e}\otimes\cdots\otimes\ket{+}_{\tau\left(\frac{N}{2}+m_N\right)}\bra{e}\otimes\ket{-}_{\tau\left(\frac{N}{2}+m_N+1\right)}\bra{e}\right.\\
&\hspace{6cm}\left.\otimes\cdots\otimes\ket{-}_{\tau(N-1)}\bra{e}\otimes\ket{-}_N\bra{e}\right)\ket{\psi_{i,N}}\vphantom{\sum_1^N}\\
\end{split}
\end{displaymath}
We can reduce this expression further by using Eq.~(\ref{final_state}) for $N-1$:
\begin{equation}
\label{page7}
\begin{split}
\left.\hat{P}_N\cdots\hat{P}_1\ket{\psi_{i,N}}\right|_{k=\left(\frac{N}{2}+m_N\right)}=&
\left(\sum_{i=1}^N\beta_{i}\chi_{iN}\right)\left.\hat{P}_{N-1}\cdots\hat{P}_1\ket{\psi_{i,N-1}}\right|_{k=\left(\frac{N}{2}+m_N-1\right)}\otimes\ket{+}\\
&\hspace{0.25cm}+\left(\sum_{i=1}^N\alpha_{i}\chi_{iN}\right)\left.\hat{P}_{N-1}\cdots\hat{P}_1\ket{\psi_{i,N-1}}\right|_{k=\left(\frac{N}{2}+m_N\right)}\otimes\ket{-}.
\end{split}
\end{equation}

According to Sect.~II, in order to generate the state $\ket{S_1,...,S_N;m_N}$ we need $k=\left(\frac{N}{2}+m_N\right)$ $\boldsymbol{\sigma}^--$polarizers, for the state $\ket{S_1,...,S_{N-1};m_N-\frac{1}{2}}$ we require $k=\left(\frac{N}{2}+m_N-1\right)$ $\boldsymbol{\sigma}^--$polarizers and for the state $\ket{S_1,...,S_{N-1};m_N+\frac{1}{2}}$ we need $k=\left(\frac{N}{2}+m_N\right)$ $\boldsymbol{\sigma}^--$polarizers. Consequently, using Eq.~(\ref{page7}), we can write the requested recursion formula:
\begin{equation}
\label{rekursion}
\begin{split}
\left.\hat{P}_N\cdots\hat{P_1}\ket{\psi_{i,N}}\right|_{\ket{S_1,...,S_N;m_N}}=&
\left(\sum_{i=1}^N\beta_{i}\chi_{iN}\right) \cdot\left.\hat{P}_{N-1}\cdots\hat{P_1}\ket{\psi_{i,N-1}}\right|_{\ket{S_1,...,S_{N-1};m_N-\frac{1}{2}}}\otimes\ket{+}\\
&+\left(\sum_{i=1}^N\alpha_{i}\chi_{iN}\right)\cdot\left.\hat{P}_{N-1}\cdots\hat{P_1}\ket{\psi_{i,N-1}}\right|_{\ket{S_1,...,S_{N-1};m_N+\frac{1}{2}}}\otimes\ket{-}.
\end{split}
\end{equation}
\end{widetext}

Next, we can compute the sums appearing in the brackets of Eq.~(\ref{rekursion}) by considering the rules given by the algorithm in Sect.~II. Hereby, we have to examine two cases:\\
\begin{enumerate}
\item $\mathbf{S_N>S_{N-1}}$\textbf{:}\\
According to the algorithm, we have to connect the $N$th atom with all detectors, if and only if these detectors weren't used for a descent $S_i<S_{i-1}$. As we have to set up exactly $\left(\frac{N}{2}+m_N\right)$ $\boldsymbol{\sigma}^--$polarizers and $\left(\frac{N}{2}-m_N\right)$ $\boldsymbol{\sigma}^+-$polarizers, there are only $\left(\frac{N}{2}+m_N\right)$ $\beta-$coefficients and $\left(\frac{N}{2}-m_N\right)$ $\alpha-$coefficients unequal to zero. These remaining coefficients as well as the dedicated phases $\chi_{jk}$ take all the values $1$. Since $k$ indicates the number of the descents, i.e., $k=\frac{N}{2}-S_N=\frac{N}{2}-\left(S_{N-1} + \frac{1}{2}\right)$, we can conclude that:
\begin{equation}
\sum_{i=1}^N \beta_i \chi_{iN}  = \left(\frac{N}{2}+m_N-k\right)=\left(S_{N-1}+m_N+\frac{1}{2}\right)
\end{equation}
\begin{equation}
\sum_{i=1}^N \alpha_i \chi_{iN} = \left(\frac{N}{2}-m_N-k\right)=\left(S_{N-1}-m_N+\frac{1}{2}\right)
\end{equation}  

\newpage

\item $\mathbf{S_N<S_{N-1}}$\textbf{:}\\
According to the algorithm we have to connect the $N-$th atom with two detectors, which are equipped with polarising filters of different orientation, i.e., one oriented along $\boldsymbol{\sigma}^-$ and the other along $\boldsymbol{\sigma}^+$. Hence, only two phase factors $\chi_{iN}$ in our sums are unequal to zero. The connection between the $N$th atom and the detector $l$ which has a $\boldsymbol{\sigma}^--$polarizer in front $(\beta_l=1 \quad \wedge\quad \alpha_l=0)$ must induce an optical phase shift of $\pi$ $(\chi_{lN}=-1)$. If we denote the other detector which is equipped with a $\boldsymbol{\sigma}^+-$polarizer in front with $m$ ($\beta_m=0 \quad \wedge \quad\alpha_m=1\quad \wedge\quad \chi_{mN}=1)$, we can conclude that:
\begin{eqnarray}
\sum_{i=1}^N \beta_i \chi_{iN} & = & \beta_l\chi_{lN}+\beta_m\chi_{mN}=-1\\
\sum_{i=1}^N \alpha_i \chi_{iN} & = & \alpha_l\chi_{lN}+\alpha_m\chi_{mN}=1
\end{eqnarray}  
\end{enumerate}

With these identities we can formulate the requested recursion formula in the following way:
\begin{widetext}
\centerline{\textbf{Recursion formula for the algorithm:}}
\begin{itemize}
\item $S_N>S_{N-1}$:\\
\begin{equation}
\label{rekursion1}
\begin{split}
&\left.\hat{P}_N\cdots\hat{P}_1\ket{\psi_{i,N}}\right|_{\ket{S_1,...,S_N;m_N}}=\\
&\hspace{1.2cm}\left(S_{N-1}+m_N+\frac{1}{2}\right)\cdot\left.\hat{P}_{N-1}\cdots\hat{P_1}\ket{\psi_{i,N-1}}\right|_{\ket{S_1,...,S_{N-1};m_N-\frac{1}{2}}}\otimes\ket{+}\\
&\hspace{1.2cm}+\left(S_{N-1}-m_N+\frac{1}{2}\right)\cdot\left.\hat{P}_{N-1}\cdots\hat{P_1}\ket{\psi_{i,N-1}}\right|_{\ket{S_1,...,S_{N-1};m_N+\frac{1}{2}}}\otimes\ket{-}.
\end{split}
\end{equation}
\item $S_N<S_{N-1}$:\\
\begin{equation}
\label{rekursion2}
\begin{split}
&\left.\hat{P}_N\cdots\hat{P_1}\ket{\psi_{i,N}}\right|_{\ket{S_1,...,S_N;m_N}}=\\
&\hspace{1.5cm}-\left.\hat{P}_{N-1}\cdots\hat{P_1}\ket{\psi_{i,N-1}}\right|_{\ket{S_1,...,S_{N-1};m_N-\frac{1}{2}}}\otimes\ket{+}\\
&\hspace{1.5cm}+\left.\hat{P}_{N-1}\cdots\hat{P_1}\ket{\psi_{i,N-1}}\right|_{\ket{S_1,...,S_{N-1};m_N+\frac{1}{2}}}\otimes\ket{-}.
\end{split}
\end{equation}
\end{itemize} 
\end{widetext}

\section{Recursivity of the coupling of angular momenta in quantum mechanics}

In this section we recapitulate the well-known procedure in quantum mechanics to obtain the eigenstates of a total angular momentum (coupled basis) from the eigenstates of the angular momenta to be coupled (bare or decoupled basis) and derive the corresponding recursion formulas. This will allow us in the next section to compare the two recursion formulas, the one from our algorithm (Eqs.~(\ref{rekursion1}) and~(\ref{rekursion2}) from Sect.~V above) and the one obtained from the quantum mechanical addition of angular momenta (see Eqs.~(\ref{add1}) and~(\ref{add2}) below), and to prove that both lead to the same result. In this way, we show that 
the algorithm presented in \cite{Maser09} and the one used in quantum mechanics to construct the eigenstates of a total angular momentum using the addition of angular momenta of $N$ spin-1/2 particles are identical.

In quantum mechanics, the total angular momentum eigenstates $\ket{J,M}$ are written in the decoupled basis $\ket{j_1,j_2;m_1,m_2}$ in the following form \cite{Cohen}
\begin{equation}
\label{allgemeine_formel}
\ket{J,M}=\sum_{m_1=-j_1}^{j_1}\sum_{m_2=-j_2}^{j_2} \braket{j_1,j_2;m_1,m_2}{J,M} \ket{j_1,j_2;m_1,m_2},
\end{equation}
where the elements of the transformation matrix $\braket{j_1,j_2;m_1,m_2}{J,M}$ are the familiar \textit{Clebsch-Gordan-Coefficients} (CGC) \cite{Cohen}. If we consider the well known selection rules
\begin{equation}
\label{auswahlregel1}
M=m_1+m_2\\
\end{equation}
\begin{equation}
\label{auswahlregel2}
\textrm{and} \quad \vert j_1-j_2 \vert \le J \le j_1+j_2.
\end{equation}
for the CGC, we can simplify Eq.~(\ref{allgemeine_formel}), since in our case $m_2$ takes only the values $\pm\frac{1}{2}$: 
\begin{equation}
\label{formel_auswahlregel1}
\begin{split}
&\ket{j_1\pm\frac{1}{2},M}=\\
&\braket{j_1,\frac{1}{2};M-\frac{1}{2},\frac{1}{2}}{j_1\pm\frac{1}{2},M}\ket{j_1,\frac{1}{2};M-\frac{1}{2},\frac{1}{2}}\\
&+\braket{j_1,\frac{1}{2};M+\frac{1}{2},-\frac{1}{2}}{j_1\pm\frac{1}{2},M}\ket{j_1,\frac{1}{2};M+\frac{1}{2},-\frac{1}{2}}
\end{split}
\end{equation}
with $J=j_1\pm\frac{1}{2}$. 
In the case $j_1=\frac{k}{2}$ with $k\in\mathbb{N}_0$ and $j_2=\frac{1}{2}$, the CGC are given by the following analytic expressions \cite{Cohen}:
\begin{equation}
\begin{split}
\braket{j_1,\frac{1}{2};M-\frac{1}{2},\frac{1}{2}}{j_1+\frac{1}{2},M}&=\sqrt{\frac{j_1+M+\frac{1}{2}}{2j_1+1}}\\
\braket{j_1,\frac{1}{2};M+\frac{1}{2},-\frac{1}{2}}{j_1+\frac{1}{2},M}&=\sqrt{\frac{j_1-M+\frac{1}{2}}{2j_1+1}}\\
\braket{j_1,\frac{1}{2};M-\frac{1}{2},\frac{1}{2}}{j_1-\frac{1}{2},M}&=-\sqrt{\frac{j_1-M+\frac{1}{2}}{2j_1+1}}\\
\braket{j_1,\frac{1}{2};M+\frac{1}{2},-\frac{1}{2}}{j_1-\frac{1}{2},M}&=\sqrt{\frac{j_1+M+\frac{1}{2}}{2j_1+1}}.
\end{split}
\end{equation}
If we denote the states $\ket{\frac{1}{2};\pm\frac{1}{2}}$ with $\ket{\pm}$ and recall that in our notation we have $j_1=S_{N-1}$ and $j_2=\frac{1}{2}$, we can write Eqs.~(\ref{formel_auswahlregel1}) also in the following form:\\

\centerline{\textbf{Recursion formula for the total angular momentum}}
\centerline{\textbf{ eigenstates with} {$\mathbf{j_1=S_{N-1}}$ \textbf{and} $\mathbf{j_2=\frac{1}{2}}$}}

\begin{itemize}
\item $\mathbf{S_{N}>S_{N-1}}$:
\begin{equation}
\label{add1}
\begin{split}
&\ket{S_1,...,S_N;m_N}=\\
&\sqrt{\frac{S_{N-1}+m_N+\frac{1}{2}}{2S_{N-1}+1}}\ket{S_1,...,S_{N-1};m_N-\frac{1}{2}}\otimes\ket{+}\\
&+\sqrt{\frac{S_{N-1}-m_N+\frac{1}{2}}{2S_{N-1}+1}}\ket{S_1,...,S_{N-1};m_N+\frac{1}{2}}\otimes\ket{-}.
\end{split}
\end{equation}
\item $\mathbf{S_N<S_{N-1}}$:
\begin{equation}
\label{add2}
\begin{split}
&\ket{S_1,...,S_N;m_N}=\\
&-\sqrt{\frac{S_{N-1}-m_N+\frac{1}{2}}{2S_{N-1}+1}}\ket{S_1,...,S_{N-1};m_N-\frac{1}{2}}\otimes\ket{+}\\
&+\sqrt{\frac{S_{N-1}+m_N+\frac{1}{2}}{2S_{N-1}+1}}\ket{S_1,...,S_{N-1};m_N+\frac{1}{2}}\otimes\ket{-}.
\end{split}
\end{equation}
\end{itemize}

\section{Proof by induction}

By using the recursion formula for our algorithm (Eqs.~(\ref{rekursion1}) and (\ref{rekursion2})) and the recursion formula for the coupled basis according to the quantum mechanical prescription for the addition of angular momenta (Eqs.~(\ref{add1}) and~(\ref{add2})), we can now demonstrate the equivalence of the two expressions and in this way prove the validity of our algorithm by mathematical induction.\\

\textbf{Proposition:}\\

The algorithm presented in \cite{Maser09} is able to generate any of the $2^N$ symmetric and nonsymmetric total angular momentum eigenstates spanning the Hilbert space of an $N$ qubit compound, that is:
\begin{equation}
\label{behauptung}
\ket{S_1,...,S_N;m_N}=A\cdot \left.\hat{P}_N\cdots\hat{P}_1\ket{\psi_{i,N}}\right|_{\ket{S_1,...,S_N;m_N}}
\end{equation}
with $A\in\mathbb{R}$.\\

\textbf{Proof:}\\

We prove this proposition with a mathematical induction for $N$.\\

\textbf{Basis:} \hspace{1cm} $N=2$\\

For $N=2$ we can directly compute the proposition. We verify the proposition by using the well known states of the \textbf{spin-1 triplet} and the state of the \textbf{spin-0 singulet} of a two qubit system \cite{Cohen} and the rules of the algorithm as formulated in Eq.~(\ref{final_state}) with $N=2$. 

As we can see in the following table, we find for every state of a two qubit system one necessary $A\in\mathbb{R}$: 
\newpage
\begin{widetext}
\begin{table}[h!]
\centering
\setlength{\tabcolsep}{20pt}
\renewcommand{\arraystretch}{1.9}
\begin{tabular}{c|c|c|c}
 & $\ket{S_1,S_2;m_N}$ & $\left.\hat{P}_2\hat{P}_1\ket{\psi_{i,2}}\right|_{\ket{S_1,S_2;m_N}}$ & $A$\\
\hline
$\ket{\frac{1}{2},1;1}$ & $\ket{++}$ & $2\ket{++}$ & $\frac{1}{2}$\\
$\ket{\frac{1}{2},1;0}$ & $\frac{1}{\sqrt{2}}\left(\ket{+-}+\ket{-+}\right)$ & $\left(\ket{+-}+\ket{-+}\right)$ & $\frac{1}{\sqrt{2}}$\\
$\ket{\frac{1}{2},1;-1}$ & $\ket{--}$ & $2\ket{--}$ & $\frac{1}{2}$\\
$\ket{\frac{1}{2},0;0}$ & $\frac{1}{\sqrt{2}}\left(\ket{+-}-\ket{-+}\right)$ & $\left(\ket{+-}-\ket{-+}\right)$ & $\frac{1}{\sqrt{2}}$
\end{tabular}
\caption{\small{\slshape{Verification of the proposition with $N=2$.}}}
\label{Induktionsanfang}
\end{table}
\end{widetext}

\textbf{Induction hypothesis:} 

Let us assume that the following identity holds for one $N\in\mathbb{N}$.
\begin{displaymath}
\begin{split}
&\ket{S_1,...,S_{N-1};m_{N-1}}=\\
&\hspace{2cm}A\cdot \left.\hat{P}_{N-1}\cdots\hat{P}_1\ket{\psi_{i,N-1}}\right|_{\ket{S_1,...,S_{N-1};m_{N-1}}} \, .
\end{split}
\end{displaymath}

\vspace{0.5cm}
\textbf{Induction step:} \hspace{1cm} $N-1\longrightarrow N$\\

Since the induction hypothesis holds for both states of the $N-1$ qubit sys\-tem: 
\begin{equation}
\label{IA1}
\begin{split}
&\ket{S_1,...,S_{N-1};m_N-\frac{1}{2}}=\\
&\hspace{2cm}A_1\cdot\left.\hat{P}_{N-1}\cdots\hat{P}_1\ket{\psi_{i,N-1}}\right|_{\ket{S_1,...,S_{N-1};m_N-\frac{1}{2}}}\\
\end{split}
\end{equation}
and
\begin{equation}
\label{IA2}
\begin{split}
&\ket{S_1,...,S_{N-1};m_N+\frac{1}{2}}=\\
&\hspace{2cm}A_2\cdot\left.\hat{P}_{N-1}\cdots\hat{P}_1\ket{\psi_{i,N-1}}\right|_{\ket{S_1,...,S_{N-1};m_N+\frac{1}{2}}}
\end{split}
\end{equation}
with $A_1,A_2\in\mathbb{R}$, we can compute recursively the state $\ket{S_1,...,S_N;m_N}$ by using Eqs.~(\ref{add1}) and~(\ref{add2}) of Sect.~VI for the quantum mechanical procedure to generate the total angular momenta eigenstates. 

In the case of $S_N>S_{N-1}$, the state is given by (see Eq.~(\ref{add1}))
\begin{displaymath}
\begin{split}
&\ket{S_1,...,S_N;m_N}=\\
&\sqrt{\frac{S_{N-1}+m_N+\frac{1}{2}}{2S_{N-1}+1}}\ket{S_1,...,S_{N-1};m_N-\frac{1}{2}}\otimes\ket{+}\\
&+\sqrt{\frac{S_{N-1}-m_N+\frac{1}{2}}{2S_{N-1}+1}}\ket{S_1,...,S_{N-1};m_N+\frac{1}{2}}\otimes\ket{-}
\end{split}
\end{displaymath}
whereas in the case of $S_N<S_{N-1}$ the state can be expressed by (see Eq.~(\ref{add2}))
\begin{displaymath}
\begin{split}
&\ket{S_1,...,S_N;m_N}=\\
&-\sqrt{\frac{S_{N-1}-m_N+\frac{1}{2}}{2S_{N-1}+1}}\ket{S_1,...,S_{N-1};m_N-\frac{1}{2}}\otimes\ket{+}\\
&+\sqrt{\frac{S_{N-1}+m_N+\frac{1}{2}}{2S_{N-1}+1}}\ket{S_1,...,S_{N-1};m_N+\frac{1}{2}}\otimes\ket{-}.
\end{split}
\end{displaymath}

If we now use the induction hypothesis (Eq.~(\ref{IA1}) and~(\ref{IA2})), we obtain for $S_N>S_{N-1}$
\begin{widetext}
\begin{equation}
\label{g1}
\begin{split}
&\ket{S_1,...,S_N;m_N}=\\
&\sqrt{\frac{S_{N-1}+m_N+\frac{1}{2}}{2S_{N-1}+1}}\cdot A_1\cdot\left.\hat{P}_{N-1}\cdots\hat{P}_1\ket{\psi_{i,N-1}}\right|_{\ket{S_1,...,S_{N-1};m_N-\frac{1}{2}}}\otimes\ket{+}\\
&+\sqrt{\frac{S_{N-1}-m_N+\frac{1}{2}}{2S_{N-1}+1}}\cdot A_2\cdot\left.\hat{P}_{N-1}\cdots\hat{P}_1\ket{\psi_{i,N-1}}\right|_{\ket{S_1,...,S_{N-1};m_N+\frac{1}{2}}}\otimes\ket{-}
\end{split}
\end{equation}
and for $S_N<S_{N-1}$
\begin{equation}
\label{g2}
\begin{split}
&\ket{S_1,...,S_N;m_N}=\\
&\hspace{1.0cm}-\sqrt{\frac{S_{N-1}-m_N+\frac{1}{2}}{2S_{N-1}+1}}\cdot A_1\cdot\left.\hat{P}_{N-1}\cdots\hat{P}_1\ket{\psi_{i,N-1}}\right|_{\ket{S_1,...,S_{N-1};m_N-\frac{1}{2}}}\otimes\ket{+}\\
&\hspace{1.5cm}+\sqrt{\frac{S_{N-1}+m_N+\frac{1}{2}}{2S_{N-1}+1}}\cdot A_2\cdot\left.\hat{P}_{N-1}\cdots\hat{P}_1\ket{\psi_{i,N-1}}\right|_{\ket{S_1,...,S_{N-1};m_N+\frac{1}{2}}}\otimes\ket{-}.
\end{split}
\end{equation}
\end{widetext}

Next, we show that the right hand sides of Eqs.~(\ref{g1}) and~(\ref{g2}) are, up to a real constant, nothing else than the right hand sides of Eqs.~(\ref{rekursion1}) and~(\ref{rekursion2}) corresponding to the recursion formula of our algorithm (see Sect.~V). This identity proves the proposition Eq.~(\ref{behauptung}). To this end, we insert Eq.~(\ref{g1}) and (\ref{g2}) and the two recursion formula of the algorithm for $S_N>S_{N-1}$ (Eq.~(\ref{rekursion1})) and for $S_N<S_{N-1}$ (Eq.~(\ref{rekursion2})) in the proposition Eq.~(\ref{behauptung}). By comparing the coefficients of the appearing states, the proposition will be reduced to the following system of equations:\\
\\
For $S_N>S_{N-1}$ the proposition reduces to 
\begin{equation}
\label{system1}
\sqrt{\frac{S_{N-1}+m_N+\frac{1}{2}}{2S_{N-1}+1}}\cdot A_1=A\left(S_{N-1}+m_N+\frac{1}{2}\right)
\end{equation}
\begin{equation}
\label{system2}
\sqrt{\frac{S_{N-1}-m_N+\frac{1}{2}}{2S_{N-1}+1}}\cdot A_2=A\left(S_{N-1}-m_N+\frac{1}{2}\right)
\end{equation}
and for $S_N<S_{N-1}$ to
\begin{equation}
\label{system3}
-\sqrt{\frac{S_{N-1}-m_N+\frac{1}{2}}{2S_{N-1}+1}}\cdot A_1=-\tilde{A}
\end{equation}
\begin{equation}
\label{system4}
\sqrt{\frac{S_{N-1}+m_N+\frac{1}{2}}{2S_{N-1}+1}}\cdot A_2=\tilde{A}.
\end{equation}
with $A, \tilde{A}\in\mathbb{R}$\\
\\
\textbf{Special case:}\\

$S_{N-1}+m_N+\frac{1}{2}=0 \quad \vee \quad S_{N-1}-m_N+\frac{1}{2}=0$\\

In this special case, one of the two CGC appearing in the quantum mechanical recursion formula for state $\ket{S_1,...,S_N;m_N}$ (Eqs.~(\ref{add1}) and~(\ref{add2})) is zero. In that case the proposition Eq.~(\ref{behauptung}) is obviously right since every equation in the form $a\cdot x=y \; (a \neq 0)$ has a unique solution in the field of the real numbers. In fact, we can easily indicate the requested $A, \tilde{A}\in\mathbb{R}$:\\

If $S_{N-1}+m_N+\frac{1}{2}=0$, $A$ is given in the case of $S_{N}>S_{N-1}$ by:
\begin{displaymath}
A=A_2\cdot\sqrt{\frac{1}{\left(2S_{N-1}+1\right)\left(S_{N-1}-m_N+\frac{1}{2}\right)}}
\end{displaymath}
whereas for the case $S_{N}<S_{N-1}$, $\tilde{A}$ is given by
\begin{displaymath}
\tilde{A}=A_1\cdot\sqrt{\frac{S_{N-1}-m_N+\frac{1}{2}}{2S_{N-1}+1}}
\end{displaymath}
If $S_{N-1}-m_N+\frac{1}{2}=0$,  $A$ is given in the case of $S_{N}>S_{N-1}$ by
\begin{displaymath}
A=A_1\cdot\sqrt{\frac{1}{\left(2S_{N-1}+1\right)\left(S_{N-1}+m_N+\frac{1}{2}\right)}}.
\end{displaymath}
whereas for $S_{N}<S_{N-1}$ $\tilde{A}$ is given by
\begin{displaymath}
\tilde{A}=A_2\cdot\sqrt{\frac{S_{N-1}+m_N+\frac{1}{2}}{2S_{N-1}+1}}.
\end{displaymath}

\textbf{General case:}\\

$S_{N-1}+m_N+\frac{1}{2}\neq 0 \quad \wedge \quad S_{N-1}-m_N+\frac{1}{2}\neq 0$\\

For all other cases, that is when all needed CGC are unequal to zero, we prove the proposition in the case of $S_N>S_{N-1}$, if (see Eqs.~(\ref{system1}) and~(\ref{system2})):
\begin{displaymath}
\alpha_1:=A_1\cdot\sqrt{\frac{1}{\left(2S_{N-1}+1\right)\left(S_{N-1}+m_N+\frac{1}{2}\right)}}
\end{displaymath}
is equal to
\begin{displaymath}
\alpha_2:=A_2\cdot\sqrt{\frac{1}{\left(2S_{N-1}+1\right)\left(S_{N-1}-m_N+\frac{1}{2}\right)}}.
\end{displaymath}
In the case of $S_N<S_{N-1}$, we prove the proposition, if (see Eqs.~(\ref{system3}) and~(\ref{system4})):
\begin{displaymath}
\tilde{\alpha}_1=A_1\cdot\sqrt{\frac{S_{N-1}-m_N+\frac{1}{2}}{2S_{N-1}+1}}
\end{displaymath}
is equal to
\begin{displaymath}
\tilde{\alpha}_2=A_2\cdot\sqrt{\frac{S_{N-1}+m_N+\frac{1}{2}}{2S_{N-1}+1}},
\end{displaymath}
i.e. in summary in all cases if
\begin{equation}
\label{frac}
\frac{A_1}{A_2}=\sqrt{\frac{S_{N-1}+m_N+\frac{1}{2}}{S_{N-1}-m_N+\frac{1}{2}}}
\end{equation}
holds.

In the following we retrieve this relation for $A_1$ and $A_2$ by employing a case differentiaton $S_{N-1}>S_{N-2}$ and $S_{N-1}<S_{N-2}$.\\

\textbf{First case: $S_{N-1}>S_{N-2}$}\\

We start to  use the recursion formula (Eq.~(\ref{add1}) and~(\ref{rekursion1})) for the states $\ket{S_1,...,S_{N-1};m_N-\frac{1}{2}}$ and $\left.\hat{P}_{N-1}\cdots\hat{P}_1\ket{\psi_{i,N-1}}\right|_{\ket{S_1,...,S_{N-1};m_N-\frac{1}{2}}}$. We thus obtain 
\begin{widetext}
\begin{equation}
\label{Fall1}
\begin{split}
\ket{S_1,...,S_{N-1};m_N-\frac{1}{2}}=
\hspace{1.5cm}\sqrt{\frac{S_{N-2}+m_N}{2S_{N-2}+1}}\ket{S_1,...,S_{N-2};m_N-1}\otimes\ket{+}\\
\hspace{1.5cm}+\sqrt{\frac{S_{N-2}-m_N+1}{2S_{N-2}+1}}\ket{S_1,...,S_{N-2};m_N}\otimes\ket{-}
\end{split}
\end{equation}
as well as
\begin{equation}
\label{Fall2}
\begin{split}
\left.\hat{P}_{N-1}\cdots\hat{P}_1\ket{\psi_{i,N-1}}\right|_{\ket{S_1,...,S_{N-1};m_N-\frac{1}{2}}}=&
\left(S_{N-2}+m_N\right)\left.\hat{P}_{N-2}\cdots\hat{P}_1\ket{\psi_{i,N-2}}\right|_{\ket{S_1,...,S_{N-2};m_N-1}}\otimes\ket{+}\\
&+\left(S_{N-2}-m_N+1\right)\left.\hat{P}_{N-2}\cdots\hat{P}_1\ket{\psi_{i,N-2}}\right|_{\ket{S_1,...,S_{N-2};m_N}}\otimes\ket{-}.
\end{split}
\end{equation}

By using the induction hypothesis (Eq.~(\ref{IA1})) we can conclude from Eqs.~(\ref{Fall1}) and~(\ref{Fall2}):
\begin{displaymath}
\begin{split}
&\sqrt{\frac{S_{N-2}+m_N}{2S_{N-2}+1}}\ket{S_1,...,S_{N-2};m_N-1}\otimes\ket{+}+\sqrt{\frac{S_{N-2}-m_N+1}{2S_{N-2}+1}}\ket{S_1,...,S_{N-2};m_N}\otimes\ket{-}=\\
&A_1\cdot\left(S_{N-2}+m_N\right)\left.\hat{P}_{N-2}\cdots\hat{P}_1\ket{\psi_{i,N-2}}\right|_{\ket{S_1,...,S_{N-2};m_N-1}}\otimes\ket{+}\\
&\hspace{3.0cm}+A_1\cdot\left(S_{N-2}-m_N+1\right)\left.\hat{P}_{N-2}\cdots\hat{P}_1\ket{\psi_{i,N-2}}\right|_{\ket{S_1,...,S_{N-2};m_N}}\otimes\ket{-}.
\end{split}
\end{displaymath}
\end{widetext}

This equation implies that the states $\ket{S_1,...,S_{N-2};m_N}$ and $\left.\hat{P}_{N-2}\cdots\hat{P}_1\ket{\psi_{i,N-2}}\right|_{\ket{S_1,...,S_{N-2};m_N}}$ are linearly dependent, i.e. it exists one $p_1\in\mathbb{R}$, so that the following equation holds:
\begin{equation}
\label{p1}
\ket{S_1,...,S_{N-2};m_N}=p_1\cdot\left.\hat{P}_{N-2}\cdots\hat{P}_1\ket{\psi_{i,N-2}}\right|_{\ket{S_1,...,S_{N-2};m_N}}
\end{equation}
with 
\begin{equation}
\label{k1}
p_1=A_1\cdot\sqrt{\left(2S_{N-2}+1\right)\left(S_{N-2}-m_N+1\right)}.
\end{equation}

In the same way we use the recursion formula for the states $\ket{S_1,...,S_{N-1};m_N+\frac{1}{2}}$ and $\left.\hat{P}_{N-1}\cdots\hat{P}_1\ket{\psi_{i,N-1}}\right|_{\ket{S_1,...,S_{N-1};m_N+\frac{1}{2}}}$ (see Eqs.~(\ref{add1}) and~(\ref{rekursion1})). Hereby, we obtain
\begin{widetext}
\begin{displaymath}
\begin{split}
\ket{S_1,...,S_{N-1};m_N+\frac{1}{2}}=&\sqrt{\frac{S_{N-2}+m_N+1}{2S_{N-2}+1}}\ket{S_1,...,S_{N-2};m_N}\otimes\ket{+}\\
&\hspace{1.5cm}+\sqrt{\frac{S_{N-2}-m_N}{2S_{N-2}+1}}\ket{S_1,...,S_{N-2};m_N+1}\otimes\ket{-}
\end{split}
\end{displaymath}
as well as
\begin{displaymath}
\begin{split}
&\left.\hat{P}_{N-1}\cdots\hat{P}_1\ket{\psi_{i,N-1}}\right|_{\ket{S_1,...,S_{N-1};m_N+\frac{1}{2}}}=\\
&\hspace{2.0cm}\left(S_{N-2}+m_N+1\right)\left.\hat{P}_{N-2}\cdots\hat{P}_1\ket{\psi_{i,N-2}}\right|_{\ket{S_1,...,S_{N-2};m_N}}\otimes\ket{+}\\
&\hspace{3.0cm}+\left(S_{N-2}-m_N\right)\left.\hat{P}_{N-2}\cdots\hat{P}_1\ket{\psi_{i,N-2}}\right|_{\ket{S_1,...,S_{N-2};m_N+1}}\otimes\ket{-}.
\end{split}
\end{displaymath}
By using the induction hypothesis (Eq.~(\ref{IA2})) we conclude that:
\begin{displaymath}
\begin{split}
&\sqrt{\frac{S_{N-2}+m_N+1}{2S_{N-2}+1}}\ket{S_1,...,S_{N-2};m_N}\otimes\ket{+}+\sqrt{\frac{S_{N-2}-m_N}{2S_{N-2}+1}}\ket{S_1,...,S_{N-2};m_N+1}\otimes\ket{-}=\\
&\hspace{1.0cm}A_2\cdot\left(S_{N-2}+m_N+1\right)\left.\hat{P}_{N-2}\cdots\hat{P}_1\ket{\psi_{i,N-2}}\right|_{\ket{S_1,...,S_{N-2};m_N}}\otimes\ket{+}\\
&\hspace{3.0cm}+A_2\cdot\left(S_{N-2}-m_N\right)\left.\hat{P}_{N-2}\cdots\hat{P}_1\ket{\psi_{i,N-2}}\right|_{\ket{S_1,...,S_{N-2};m_N+1}}\otimes\ket{-}.
\end{split}
\end{displaymath}
\end{widetext}
This equation again implies that the states $\ket{S_1,...,S_{N-2};m_N}$ and $\left.\hat{P}_{N-1}\cdots\hat{P}_1\ket{\psi_{i,N-2}}\right|_{\ket{S_1,...,S_{N-2};m_N}}$ are linearly dependent, i.e. it exists one $p_2\in\mathbb{R}$, for which holds:
\begin{equation}
\label{p2}
\ket{S_1,...,S_{N-2};m_N}=p_2\cdot\left.\hat{P}_{N-2}\cdots\hat{P}_1\ket{\psi_{i,N-2}}\right|_{\ket{S_1,...,S_{N-2};m_N}}
\end{equation}
with
\begin{equation}
\label{k2}
p_2=A_2\cdot\sqrt{\left(2S_{N-2}+1\right)\left(S_{N-2}+m_N+1\right)}.
\end{equation}

Note that we can write the state $\ket{S_1,...,S_{N-2};m_N}$ always as a linear combination of the decoupled basis ${\ket{u_i}}$ ($i \in\{1,...,2^{N-2}\}$):
\begin{displaymath}
\ket{S_1,...,S_{N-2};m_N}=\sum_{i=1}^{2^{N-2}} c_i \ket{u_i}
\end{displaymath}
with $c_i\in\mathbb{R}$. Beyond that, we can also write the state $\left.\hat{P}_{N-2}\cdots\hat{P}_1\ket{\psi_{i,N-2}}\right|_{\ket{S_1,...,S_{N-2};m_N}}$ as a linear combination of the bare basis:
\begin{displaymath}
\left.\hat{P}_{N-2}\cdots\hat{P}_1\ket{\psi_{i,N-2}}\right|_{\ket{S_1,...,S_{N-2};m_N}}=\sum_{i=1}^{2^{N-2}} d_i \ket{u_i}
\end{displaymath}
with $d_i\in\mathbb{R}$. As we can write one state in the Hilbert space for a given basis in just one unique way, we can conclude that due to Eqs.~(\ref{p1}) and~(\ref{p2}) we have:
\begin{displaymath}
p_1=\frac{c_i}{d_i}=p_2.
\end{displaymath}

Hence, we obtain by using Eq.~(\ref{k1}) and~(\ref{k2}):
\begin{displaymath}
\begin{split}
A_1\cdot&\sqrt{\left(2S_{N-2}+1\right)\left(S_{N-2}-m_N+1\right)}=\\
&A_2\cdot\sqrt{\left(2S_{N-2}+1\right)\left(S_{N-2}+m_N+1\right)}
\end{split}
\end{displaymath} 
and thus
\begin{equation}
\frac{A_1}{A_2}=\sqrt{\frac{S_{N-2}+m_N+1}{S_{N-2}-m_N+1}}.
\end{equation}

As we assumed in this case $S_{N-1}>S_{N-2}$, i.e.
\begin{displaymath}
S_{N-2}=S_{N-1}-\frac{1}{2},
\end{displaymath}
we have proven the equation  
\begin{equation}
\tag{\ref{frac}}
\frac{A_1}{A_2}=\sqrt{\frac{S_{N-1}+m_N+\frac{1}{2}}{S_{N-1}-m_N+\frac{1}{2}}}.
\end{equation}
for this case.\\

\textbf{Second case: $S_{N-1}<S_{N-2}$}\\

We can handle this case in the same way we treated the case $S_{N-1}>S_{N-2}$. By using the recursion formula for the states $\ket{S_1,...,S_{N-1};m_N-\frac{1}{2}}$ and $\left.\hat{P}_{N-1}\cdots\hat{P}_1\ket{\psi_{i,N-1}}\right|_{\ket{S_1,...,S_{N-1};m_N-\frac{1}{2}}}$ for $S_{N-1}<S_{N-2}$ (see Eqs.~(\ref{add2}) and~(\ref{rekursion2})), we obtain:
\begin{displaymath}
\begin{split}
&\ket{S_1,...,S_{N-1};m_N-\frac{1}{2}}=\\
&\hspace{0.5cm}-\sqrt{\frac{S_{N-2}-m_N+1}{2S_{N-2}+1}}\ket{S_1,...,S_{N-2};m_N-1}\otimes\ket{+}\\
&\hspace{1.0cm}+\sqrt{\frac{S_{N-2}+m_N}{2S_{N-2}+1}}\ket{S_1,...,S_{N-2};m_N}\otimes\ket{-}
\end{split}
\end{displaymath}
as well as
\begin{widetext}
\begin{displaymath}
\begin{split}
\left.\hat{P}_{N-1}\cdots\hat{P}\ket{\psi_{i,N-1}}\right|_{\ket{S_1,...,S_{N-1};m_N-\frac{1}{2}}}=&-\left.\hat{P}_{N-2}\cdots\hat{P}_1\ket{\psi_{i,N-2}}\right|_{\ket{S_1,...,S_{N-2};m_N-1}}\otimes\ket{+}\\
&\hspace{2cm}+\left.\hat{P}_{N-2}\cdots\hat{P}_1\ket{\psi_{i,N-2}}\right|_{\ket{S_1,...,S_{N-2};m_N}}\otimes\ket{-}.
\end{split}
\end{displaymath}

By means of the induction hypothesis Eq.~(\ref{IA1}) we can thus conclude that:
\begin{displaymath}
\begin{split}
&-\sqrt{\frac{S_{N-2}-m_N+1}{2S_{N-2}+1}}\ket{S_1,...,S_{N-2};m_N-1}\otimes\ket{+}+\sqrt{\frac{S_{N-2}+m_N}{2S_{N-2}+1}}\ket{S_1,...,S_{N-2};m_N}\otimes\ket{-}=\\
&\hspace{1.5cm}-A_1 \cdot \left.\hat{P}_{N-2}\cdots\hat{P}_1\ket{\psi_{i,N-2}}\right|_{\ket{S_1,...,S_{N-2};m_N-1}}\otimes\ket{+}+A_1 \cdot  \left.\hat{P}_{N-2}\cdots\hat{P}_1\ket{\psi_{i,N-2}}\right|_{\ket{S_1,...,S_{N-2};m_N}}\otimes\ket{-}.
\end{split}
\end{displaymath}
\end{widetext}
This equation implies that the states $\ket{S_1,...,S_{N-2};m_N}$ and $\left.\hat{P}_{N-2}\cdots\hat{P}_1\ket{\psi_{i,N-2}}\right|_{\ket{S_1,...,S_{N-2};m_N}}$ are linearly dependent, i.e. it exists one $p_1\in\mathbb{R}$, for which the following equation holds:
\begin{equation}
\label{q1}
\ket{S_1,...,S_{N-2};m_N}=p_1\cdot\left.\hat{P}_{N-2}\cdots\hat{P}_1\ket{\psi_{i,N-2}}\right|_{\ket{S_1,...,S_{N-2};m_N}}
\end{equation}
with
\begin{equation}
\label{qk1}
p_1=A_1\cdot\sqrt{\frac{2S_{N-2}+1}{S_{N-2}+m_N}}.
\end{equation}

Using again the recursion formula for the states $\ket{S_1,...,S_{N-1};m_N+\frac{1}{2}}$ and $\left.\hat{P}_{N-1}\cdots\hat{P}_1\ket{\psi_{i,N-1}}\right|_{\ket{S_1,...,S_{N-1};m_N+\frac{1}{2}}}$ in the case of $S_{N-1}<S_{N-2}$ (see Eqs.~(\ref{add2}) and~(\ref{rekursion2})), we obtain:

\begin{displaymath}
\begin{split}
&\ket{S_1,...,S_{N-1};m_N+\frac{1}{2}}=\\
&\hspace{0.5cm}-\sqrt{\frac{S_{N-2}-m_N}{2S_{N-2}+1}}\ket{S_1,...,S_{N-2};m_N}\otimes\ket{+}\\
&\hspace{1cm}+\sqrt{\frac{S_{N-2}+m_N+1}{2S_{N-2}+1}}\ket{S_1,...,S_{N-2};m_N+1}\otimes\ket{-}
\end{split}
\end{displaymath}
as well as
\begin{widetext}
\begin{displaymath}
\begin{split}
\left.\hat{P}_{N-1}\cdots\hat{P}\ket{\psi_{i,N-1}}\right|_{\ket{S_1,...,S_{N-1};m_N+\frac{1}{2}}}=&
-\left.\hat{P}_{N-2}\cdots\hat{P}_1\ket{\psi_{i,N-2}}\right|_{\ket{S_1,...,S_{N-2};m_N}}\otimes\ket{+}\\
&\hspace{2cm}+\left.\hat{P}_{N-2}\cdots\hat{P}_1\ket{\psi_{i,N-2}}\right|_{\ket{S_1,...,S_{N-2};m_N+1}}\otimes\ket{-}.
\end{split}
\end{displaymath}

By means of the induction hypothesis Eq.~(\ref{IA2}) we know that
\begin{displaymath}
\begin{split}
&-\sqrt{\frac{S_{N-2}-m_N}{2S_{N-2}+1}}\ket{S_1,...,S_{N-2};m_N}\otimes\ket{+}
+\sqrt{\frac{S_{N-2}+m_N+1}{2S_{N-2}+1}}\ket{S_1,...,S_{N-2};m_N+1}\otimes\ket{-}=\\
&\hspace{2.5cm}-A_2 \cdot \left.\hat{P}_{N-2}\cdots\hat{P}_1\ket{\psi_{i,N-2}}\right|_{\ket{S_1,...,S_{N-2};m_N}}\otimes\ket{+}
+A_2 \cdot \left.\hat{P}_{N-2}\cdots\hat{P}_1\ket{\psi_{i,N-2}}\right|_{\ket{S_1,...,S_{N-2};m_N+1}}\otimes\ket{-}.
\end{split}
\end{displaymath}
holds.
\end{widetext}

Hereby, it is again implied that the states $\ket{S_1,...,S_{N-2};m_N}$ and $\left.\hat{P}_{N-2}\cdots\hat{P}_1\ket{\psi_{i,N-2}}\right|_{\ket{S_1,...,S_{N-2};m_N}}$ are linearly dependent, i.e. it exists one $p_2\in\mathbb{R}$, for which holds:
\begin{equation}
\label{q2}
\ket{S_1,...,S_{N-2};m_N}=p_2\cdot\left.\hat{P}_{N-2}\cdots\hat{P}_1\ket{\psi_{i,N-2}}\right|_{\ket{S_1,...,S_{N-2};m_N}}
\end{equation}
with 
\begin{equation}
\label{qk2}
p_2=A_2\cdot\sqrt{\frac{2S_{N-2}+1}{S_{N-2}-m_N}}.
\end{equation}
As we can write one state in the Hilbert space for a given basis just in one unique way, we can conclude in the same way as in the first case that:
\begin{displaymath}
p_1=p_2.
\end{displaymath}
Therewith, we obtain by means of Eqs.~(\ref{qk1}) and~(\ref{qk2}):
\begin{displaymath}
\frac{A_1}{A_2}=\sqrt{\frac{S_{N-2}+m_N}{S_{N-2}-m_N}} \,.
\end{displaymath}
As we consider the case $S_{N-1}<S_{N-2}$, i.e.
\begin{displaymath}
S_{N-2}=S_{N-1}+\frac{1}{2},
\end{displaymath}
we have again proven the equation
\begin{equation}
\tag{\ref{frac}}
\frac{A_1}{A_2}=\sqrt{\frac{S_{N-1}+m_N+\frac{1}{2}}{S_{N-1}-m_N+\frac{1}{2}}}.
\end{equation}

\vspace{0.5cm}
The proven constraint for $A_1$ and $A_2$ (Eq.~(\ref{frac})) is sufficient for the conclusion that $\alpha_1=\alpha_2$ as well as $\tilde{\alpha_1}=\tilde{\alpha_2}$ holds. This means that altogether we can conclude that:
\begin{widetext}
\begin{displaymath}
\begin{split}
\alpha_1&=A_1\cdot\sqrt{\frac{1}{\left(2S_{N-1}+1\right)\left(S_{N-1}+m_N+\frac{1}{2}\right)}}=
\underbrace{A_2\sqrt{\frac{S_{N-1}+m_N+\frac{1}{2}}{S_{N-1}-m_N+\frac{1}{2}}}}_{=A_1 \quad\textrm{by using Eq.~(\ref{frac})}}\cdot\sqrt{\frac{1}{\left(2S_{N-1}+1\right)\left(S_{N-1}+m_N+\frac{1}{2}\right)}}=\\
&=A_2\cdot\sqrt{\frac{1}{\left(2S_{N-1}+1\right)\left(S_{N-1}-m_N+\frac{1}{2}\right)}}=\alpha_2
\end{split}
\end{displaymath}
\end{widetext}
as well as that:
\begin{displaymath}
\begin{split}
\tilde{\alpha_1}&=A_1\cdot\sqrt{\frac{S_{N-1}-m_N+\frac{1}{2}}{2S_{N-1}+1}}=\\
&=\underbrace{A_2\cdot\sqrt{\frac{S_{N-1}+m_N+\frac{1}{2}}{S_{N-1}-m_N+\frac{1}{2}}}}_{=A_1\quad 
\textrm{by using Eq.~(\ref{frac})}}\cdot\sqrt{\frac{S_{N-1}-m_N+\frac{1}{2}}{2S_{N-1}+1}}=\\
&=A_2\cdot\sqrt{\frac{S_{N-1}+m_N+\frac{1}{2}}{2S_{N-1}+1}}=\tilde{\alpha_2}.
\end{split}
\end{displaymath}
Hence, the proposition is proven with $A=\alpha_1=\alpha_2$ in the case of $S_N>S_{N-1}$ and with $A=\tilde{\alpha_1}=\tilde{\alpha_2}$ in the case of $S_N<S_{N-1}$, respectively.\\

\section{Conclusion and Outlook}

In this paper we have proven the algorithm introduced in \cite{Maser09} enabling the generation of any of the $2^N$ symmetric and nonsymmetric total angular momentum eigenstates spanning the Hilbert space of an $N$ qubit compound. The algorithm is implemented in a system of $N$ single photon emitters with a $\Lambda$-configuration, scattering photons from an excited state $\ket{e}$ and decaying into ground states $\ket{+}$ and $\ket{-}$, which form the individual qubit. Hereby, it is assumed that the $N$ photons are recorded by $N$ detectors such that the individual source of a registered photon remains unknown. For the proof, we formulated the algorithm in form of a  recursion formula what allowed us to compare it to the well-known quantum mechanical procedure for the generation of the eigenstates of a total angular momentum obtained by coupling a spin ${\bf S}_{N-1}$ with the spin of a further qubit. By comparing the two recursion formulas and demonstrating their equivalence, we verified the general validity of our scheme. A particular feature of the algorithm is that it allows to encode the total spin eigenstates among independent and potentially far distant qubits. This means that the method  is able to implement the required CGCs, needed to form those states, even though the qubits never directly interact with each other. In this way it simulates the coupling of the angular momenta of $N$ non-interacting qubits.

Note that in \cite{Maser10} we presented a scheme for the generation of all symmetric and non-symmetric total spin eigenstates encoded in the polarization degrees of freedom of $N$ photons. In the paper it was shown, that the methods for the generation of the polarization entangled photon states and for the generation of the entangled matter qubit states are mathematically equivalent. Thus, the proof of the algorithm given in this paper proves as well the algorithm derived in \cite{Maser10} for the generation of all symmetric and non-symmetric total spin eigenstates encoded in the polarization degrees of freedom of $N$ photons.

\section{Acknowledgments}

U.S. thank the Elite Network of Bavaria for financial support. A. M., U.S. and J. v. Z. gratefully acknowledge funding by the Erlangen Graduate School in Advanced Optical Technologies (SAOT) by the German National Science Foundation (DFG) in the framework of the German excellence initiative. C. A., A. M., U.S. and J. v. Z. thank the German National Science Foundation (DFG) and T.B. the Belgian F.R.S.-FNRS for supporting this project.

\end{document}